\documentclass{article}
\usepackage{spconf,amsmath,graphicx}

\usepackage{multirow}
\usepackage{diagbox}
\usepackage{amsmath,amssymb,amsfonts}
\usepackage{amsmath}
\usepackage{mathrsfs}

\usepackage{algorithmic}
\usepackage{graphicx}
\usepackage{textcomp}
\usepackage[table,xcdraw]{xcolor}
\usepackage{ragged2e}
\usepackage{bbm}
\usepackage{multirow}
\usepackage{tabularray}

\usepackage[style=ieee,maxcitenames=2, mincitenames=1, uniquelist=false,backend=bibtex]{biblatex}
\renewcommand*{\bibfont}{\small}
\usepackage[linesnumbered,ruled,vlined]{algorithm2e}
\usepackage[caption=false, font=footnotesize]{subfig}
\usepackage{mathtools} 
\usepackage[hang,flushmargin]{footmisc}
\usepackage{lipsum}

\usepackage{adjustbox}
\usepackage{caption}
\usepackage{enumitem}
\usepackage{colortbl}

\makeatletter
\newcommand{\algorithmfootnote}[2][\footnotesize]{%
  \let\old@algocf@finish\@algocf@finish%
  \def\@algocf@finish{\old@algocf@finish%
    \leavevmode\rlap{\begin{minipage}{\linewidth}
    #1#2
    \end{minipage}}%
  }%
}
\makeatother
\usepackage[british]{babel}
\usepackage[autostyle]{csquotes}
\usepackage{placeins}
\usepackage{float}

\def\BibTeX{{\rm B\kern-.05em{\sc i\kern-.025em b}\kern-.08em
    T\kern-.1667em\lower.7ex\hbox{E}\kern-.125emX}}
\usepackage[colorlinks,urlcolor=blue,linkcolor=blue,citecolor=blue]{hyperref}
\usepackage{caption}
\def\x{{\mathbf x}}
\def\L{{\cal L}}

\title{Syn-Att: Synthetic Speech Attribution via Semi-Supervised \\ Unknown Multi-Class Ensemble of CNNs}

\name{
\begin{tabular}{c}
Md Awsafur Rahman${}^{\text{\textsection}, 1}$, Bishmoy Paul ${}^{\text{\textsection}, 1}$, Najibul Haque Sarker ${}^{\text{\textsection}, 2}$, Zaber Ibn Abdul Hakim ${}^{\text{\textsection}, 2}$\\
Shaikh Anowarul Fattah ${}^{1}$ and  Mohammad Saquib  ${}^{3}$\end{tabular}}
\address{${}^1$ Dept. of EEE, BUET, Bangladesh\\
         ${}^2$ Dept. of CSE, BUET, Bangladesh \\ 
         ${}^3$ Dept. of EE, UT Dallas, Texas, USA }

\begin{document}

\maketitle

\begingroup\renewcommand\thefootnote{\textsection}
\footnotetext{Equal contribution}
\endgroup

\begin{abstract}
With the huge technological advances introduced by deep learning in audio \& speech processing, many novel synthetic speech techniques achieved incredible realistic results. As these methods generate realistic fake human voices, they can be used in malicious acts such as people imitation, fake news, spreading, spoofing, media manipulations, etc. Hence, the ability to detect synthetic or natural speech has become an urgent necessity. Moreover, being able to tell which algorithm has been used to generate a synthetic speech track can be of preeminent importance to track down the culprit. In this paper, a novel strategy is proposed to attribute a synthetic speech track to the generator that is used to synthesize it. The proposed detector transforms the audio into log-mel spectrogram, extracts features using CNN, and classifies it between five known and unknown algorithms, utilizing semi-supervision and ensemble to improve its robustness and generalizability significantly.  The proposed detector is validated on two evaluation datasets consisting of a total of 18,000 weakly perturbed (Eval 1) \& 10,000 strongly perturbed (Eval 2) synthetic speeches. The proposed method\footnote{Code \& Dataset is available at \href{https://github.com/awsaf49/synatt}{https://github.com/awsaf49/synatt}} outperforms other top teams in accuracy by 12-13\% on Eval 2 and 1-2\% on Eval 1, in the IEEE SP Cup challenge at ICASSP 2022.
\end{abstract}


\section{Introduction and Related Work}
Due to the utilization of audio in tasks related to security, privacy, evidence and more non-frivolous activities, the quantitative and qualitative research in audio and its discipline has surged in recent times. With the advent of deep learning technologies, an array of new methods has been introduced for voice and speech recognition and comprehension~\cite{syn-speech-att, data-driven}. This improvement in technology also is evident in the synthetic speech generation field which is now in such a state that even synthetic speech of an individual can be mimicked flawlessly~\cite{syn1, data-driven, transformer}. This has given rise to the possibility that the technology can now be used for malevolent purposes and poses security concerns which cannot be ignored~\cite{mediaforensics}.

In order to combat the proliferation of illegal and detrimental activities, the development of technologies for the detection and classification of fake speeches~\cite{speaker_tamper} is of paramount importance, not only from a law enforcement perspective but also within the context of machine ethics. While numerous efforts have already been made in the field of forensic detectors designed to differentiate between genuine speech recordings and synthetically generated ones~\cite{fake_audio}, the challenge of attributing a synthetic speech track to the specific generator used for its synthesis remains relatively unexplored. Traditional methods, relying on closed-set approaches such as simple classification~\cite{xgboost, convnext, wollmer2013lstm}, and consistency detection~\cite{an2015variational}, fall short in detecting samples from unseen algorithms (open-set scenarios). These methods tend to confuse samples from unknown algorithms with known ones, resulting in subpar performance. Recent attempts, such as ParalMGC~\cite{neri2022paralmgc}, have aimed to address the issue of unknown algorithms by employing parallel branches (utilizing Mel-Frequency and GammaTone coefficients) CNNs but it struggles when faced with unseen perturbed test cases. Another method, CAT~\cite{transformer}, leverages transformers in conjunction with t-SNE to identify unknown algorithms based on latent space but falters when confronted with substantial variations within known algorithm due to factors like speaker changes or environmental shifts. An alternative data-driven approach~\cite{data-driven} has specifically targeted the challenge of addressing unknown algorithms with a confidence threshold and a one-class SVM. While both of these methods exhibit promising results, they suffer from a lack of robustness, due to their reliance on highly perturbable confidence parameter, resulting in poor performance in strongly perturbed cases. To mitigate the aforementioned challenges, a novel approach is proposed which exploits a multi-class strategy with semi-supervision and ensemble techniques to attribute both known and unknown synthetic speech algorithms, ensuring robustness and generalizability.

\vspace*{-0.4cm}
\section{Methodology}

\subsection{Problem Formulation}
Mathematically, given a data set $S = \{(x_1, Y_1),\cdots , (x_N , Y_N )\}$ where N is the number of sample, $x_i$ denotes the $i^{th}$ audio sample and $Y_i$ represents the $i^{th}$ label.
The experimented approaches can majorly be classified in two classes. Firstly, using raw audio as the input feature and secondly, using log-mel spectrogram instead. \ 
If, $T(\cdot)$ denotes the transformation that extracted log-mel spectrogram from raw audio and $F(\cdot)$ denotes any generic method that generated prediction label, $\hat{y}$, from feature, then
\begin{equation*}
    \hat{y} = F(x_i) \quad \text{or} \quad \hat{y} = F(T(x_i))  
\end{equation*}
Following this, total loss, $g$, was calculated using a loss function.
\begin{equation*}
    g = \sum_{i=1}^{N} Loss(\hat{y_i}, Y_i)
\end{equation*}
The target was to minimize $g$.

\subsection{Unknown Multi-Class Strategy}
\begin{figure}[h]
    \centering
    \includegraphics[scale=0.09]{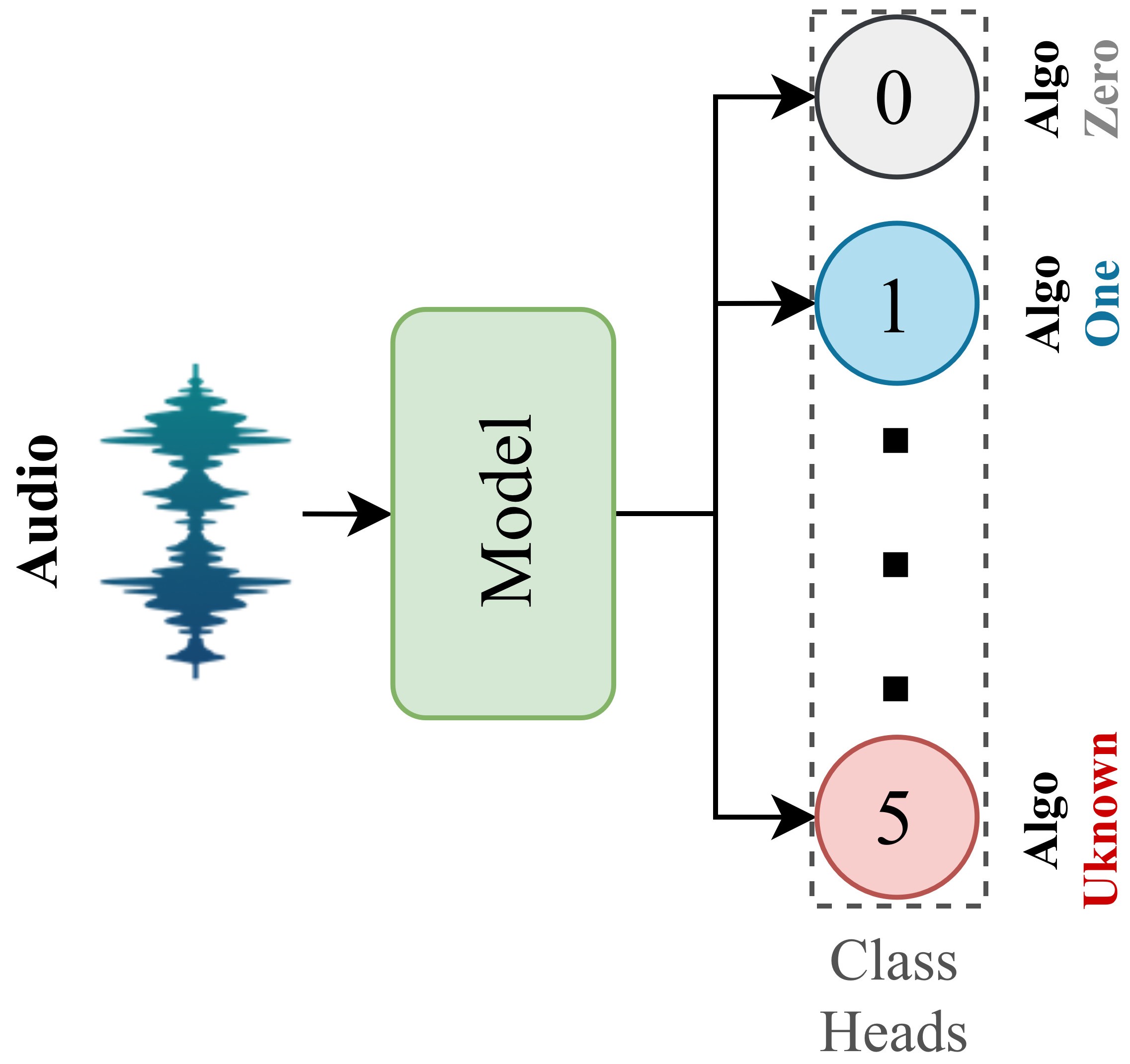}
    \caption{Proposed Unknown Multi-Class Scheme}
    \label{fig:model}
\end{figure}

To identify unknown algorithms an additional class has been added namely "Unknown" class along with data of 5 classes. The data for this additional class is added in both training and validation phase to make the distribution as diverse as possible with help of external data. Thus, synthetic speech attribution from both known and unknown class has been formulated as a Six Class Classification problem. Fig.~\ref{fig:model} provides a visual insight on the proposed unknown multi-class scheme.

\subsection{Data Processing}
In the data pipeline, all input signals are resampled to 16,000 samples/second and then Z-normalized. Subsequently, log-mel spectrograms are generated as model inputs. In Part I, Random 6-second segments are extracted from audio signals, and shorter ones are randomly padded. The resulting log-mel spectrograms are 128 x 384, with a hop length of 250, 128 mel bins, and FFT/window sizes of 2048. In Part II, 8-second sequences are utilized for improved noise handling, with all parameters remaining the same, except for an increase in mel bins to 256, resulting in spectrograms of 256 x 512. For evaluation, two datasets are available: Eval 1 and Eval 2. Eval 2 contains strong perturbations (pitch shift, time stretch, filtering), making it very challenging. Eval 1 consists of two parts—one without perturbations and one with weak perturbations (noise, compression, reverberation). The final Eval 1 result is computed as $0.7 \times \text{Part I} + 0.3 \times \text{Part II}$ to balance contributions.

For training, 1000 samples per algorithm (0, 1, 2, 3, 4) are provided, along with an additional 1000 samples from an unseen algorithm (considered as class 5 as per proposed multi-class strategy). Classes 1, 2, 3, and 5 share a common speaker, while class 0 has a distinct speaker, and class 4 involves multiple speakers. Three publicly available natural speech datasets (LJSpeech~\cite{ljspeech17}, LibriSpeech~\cite{panayotov2015librispeech}, VCTK~\cite{veaux2017superseded}) are included as unseen algorithms, aiming to 1) diversify the unknown class, 2) mitigate speaker-specific overfitting, and 3) enhance generalization. It's important to highlight that the Eval data does not contain natural speech, allowing for the inclusion of natural speech in the unknown class. Additionally, if necessary, distinguishing between predicted natural speech and synthetic speech can be easily accomplished using conventional methods. To further diversify the unknown class, synthetic data is generated through various algorithms. Texts are extracted from 5000 training samples, utilizing the Wav2Vec 2.0 model~\cite{baevski2020wav2vec} for initial extraction, correcting spelling inconsistencies with NeuralSpeechCorrector~\cite{jayanthi2020neuspell}, and then processing the text with various text-to-speech models~\cite{tatotron2, fastpitch, fastspeech} to produce synthetic audio.

\subsection{Ensemble}
In order to improve the comprehensive representation of multifaceted features inherent in the input dataset, an ensemble methodology is strategically employed. This approach harmoniously amalgamates the outcomes of individual models, culminating in a cohesive, resilient, and universally applicable result. This ensemble strategy is characterized by the utilization of the mean operation applied to the probability outputs from multiple models, thereby yielding the ultimate prediction..

\subsection{Semi-Supervised Training}
\begin{figure}[h]
    \centering
    \includegraphics[scale=0.35]{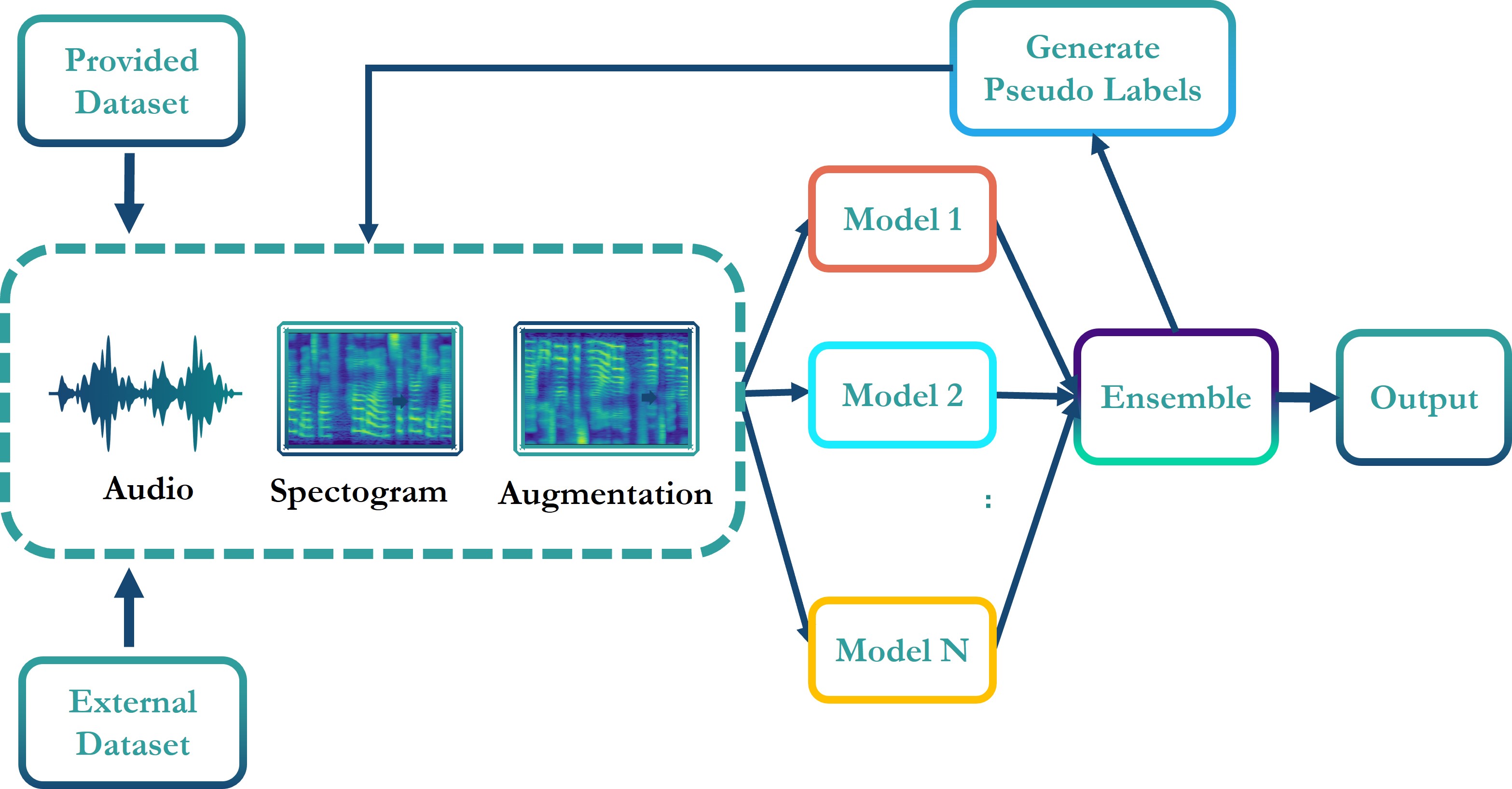}
    \caption{Proposed Semi-Supervised Scheme}
    \label{fig:solution_pipeline}
\end{figure}

The proposed approach leverages Semi-Supervised Training~\cite{pseudo}, commonly referred to as Pseudo Labelling, to enhance model robustness and generalizability. This technique involves generating approximate labels for input data based on the features learned during training. Initially, a model is trained using both provided and external datasets, followed by soft label (no thresholding) generation on the test set. These generated labels, termed pseudo labels, are not guaranteed to be ground truth and may exhibit bias towards training labels. However, by incorporating these pseudo labels as additional training data, the model adapts to the test data distribution, resulting in improved learning sample space. For a visual representation of the approach, refer to Fig. \ref{fig:solution_pipeline}.

\section{Results and Discussions}

\subsection{Experimental Setup}
The hardware configuration includes $8$ cores CPU, $64$ GB RAM, and $4\times$ NVIDIA V $100$ GPUs. Various hyperparameters are selected through experimentation such as Adam optimizer, a fixed learning rate ($\gamma_1=10^{-3}$) and an ExponentialDecay scheduler in both \textbf{Part I} and \textbf{Part II}. Categorical Cross Entropy loss is used to optimize the six class CNN classifiers with label smoothing ($\alpha$ = 0.05). Diverse networks are trained with varying epochs and batch sizes for enhanced performance, incorporating a five-fold cross-validation scheme for robust validation. Model performance evaluation favors the $F1$ Score metric to tackle class imbalance introduced by external datasets.
Augmentation techniques, such as MixUp~\cite{mixup}, CutMix~\cite{cutmix}, GaussianNoise, Time-Freq Mask, JpegCompress, Crop, Pad, and others, are applied to improve robustness.

\subsection{Ablation Study}

An overview of the quantitative comparison of various stages within the ablation study is presented in Table~\ref{table:stage}, shedding light on the significance of different aspects of the proposed method. Evidently, unknown mutli-class strategy emerges as the most influential factor, given its crucial role in identifying unknown algorithms. Moreover, for single-model, semi-supervised approach surpasses augmentation methods, due to its exceptional adaptability to unknown distribution.
\begin{table}
\caption{Scores of different stages of ablation study}
\resizebox{8cm}{!}{
\label{table:stage}
\begin{tabular}{|c|c|c|c|c|} 
\hline
\multirow{2}{*}{\diagbox{Stage}{Score}} & \multicolumn{2}{c|}{Part1} & \multicolumn{2}{c|}{Part2} \\ 
\cline{2-5}
 & CV (F1) & LB (Acc) & CV (F1) & LB (Acc) \\ 
\hline
Baseline & 0.926 & 0.915 & 0.919 & 0.903 \\ 
\hline
Unknown Multi-class & 0.962 & 0.948 & 0.937 & 0.934 \\ 
\hline
Data Augmentation & 0.935 & 0.929 & 0.94 & 0.926 \\ 
\hline
Ensemble & 0.944 & 0.932 & 0.942 & 0.921 \\ 
\hline
Semi-Supervised & 0.932 & 0.934 & 0.933 & 0.918 \\ 
\hline
\end{tabular}
}
\end{table}

\subsubsection{Effect of unknown multi-class strategy}

The effect of proposed multi-class strategy has been examined with respect to the variation of external datasets and backbones. The efficacy of the multi-class method is contingent upon the diversity of the unknown class. Since the provided datasets lack the diversity, external datasets have been incorporated. As Table~\ref{table:external} (w/o ensemble, augment, semi-sup.) reveals, the optimal outcome arises from the integration of distinct datasets, surpassing the baseline method by an approximate margin of $4\%$. The verification of the multi-class strategy's efficacy is further carried out through testing with various CNN backbones, as delineated in Table~\ref{table:backbone} (w/ augment, and semi-sup.), affirming the method's effectiveness across diverse backbones. Notably, in \textbf{Part I} (w/o perturbed), smaller models dominates, showcasing their resilience to overfitting attributed to their compact size. Conversely, in \textbf{Part II} (w/ perturbed), larger models gains superiority, leveraging their large complexity to adeptly extract intricate features.

\begin{table}
\caption{Effect of Multi-Class Strategy w.r.t Datasets}
\resizebox{8cm}{!}{
\label{table:external}
\begin{tabular}{|c|c|c|c|c|} 
\hline
\multirow{2}{*}{\diagbox{Dataset}{Score}} & \multicolumn{2}{c|}{Part1} & \multicolumn{2}{c|}{Part2} \\ 
\cline{2-5}
 & CV (F1) & LB (Acc) & CV (F1) & LB (Acc) \\ 
\hline
Baseline & 0.926 & 0.915 & 0.919 & 0.903 \\ 
\hline
LJSpeech~\cite{ljspeech17} & 0.937 & 0.923 & 0.928 & 0.916 \\ 
\hline
VCTK~\cite{veaux2017superseded} & 0.935 & 0.928 & 0.924 & 0.919 \\ 
\hline
LibriSpeech~\cite{panayotov2015librispeech} & 0.940 & 0.931 & 0.902 & 0.873 \\ 
\hline
Synthetic & 0.942 & 0.935 & 0.930 & 0.922 \\ 
\hline
\textbf{Best} & \textbf{0.962 }& \textbf{0.948} & \textbf{0.937} & \textbf{0.934} \\ 
\hline
\end{tabular}
}
\end{table}

\begin{table}
\centering
\caption{Effect of Multi-Class Strategy w.r.t Backbones}
\resizebox{8cm}{!}{
\label{table:backbone}
\begin{tabular}{|c|c|c|c|c|} 
\hline
\multirow{2}{*}{\diagbox{Backbone}{Score}} & \multicolumn{2}{c|}{Part1} & \multicolumn{2}{c|}{Part2} \\ 
\cline{2-5}
 & CV (F1)  & LB (Acc) & CV (F1) & LB (Acc)  \\
 \hline
ResNet50D~\cite{resnet-rs}                                          & 0.963     & 0.949          & 0.926     & 0.920           \\
\hline
ResNest50~\cite{resnet-rs} & 0.960 & 0.952 & 0.933 & 0.927 \\
\hline
ResNetRS50~\cite{resnet-rs}                                        & 0.956     & 0.955          & 0.929     & 0.918           \\
\hline
EfficientNetV2S~\cite{effnet} & 0.964 & 0.959 & 0.935 & 0.931 \\
\hline
RegNetZD8~\cite{resnet-rs}                                          & 0.969     & 0.951          & 0.941     & 0.946           \\
\hline
EfficientNetB0~\cite{effnet} & 0.971 & 0.962 & 0.958 & 0.957 \\ 
\hline
ECA\_NFNetL2~\cite{nfnet} & 0.948 & 0.930 & 0.955 & 0.949 \\ 
\hline
ConvNeXt\_Base\_22k~\cite{convnext} & 0.933 & 0.932 & 0.949 & 0.948 \\ 
\hline
ConvNeXt\_large\_22k~\cite{convnext} & 0.941 & 0.929 & 0.952 & 0.950 \\ 
\hline
ResNetRS152~\cite{resnet-rs} & 0.936 & 0.927 & 0.957 & 0.949 \\ 
\hline
EfficientNetV2M~\cite{effnet} & 0.930 & 0.922 & 0.955 & 0.952 \\
\hline
\end{tabular}
}
\end{table}

\subsubsection{Effect of different augmentations}

To combat speaker bias, Mixup~\cite{mixup} and Cutmix~\cite{cutmix} is employed. Random beta distribution ($\alpha=2.5$, $\beta=2.5$) is used to determine sample contributions. Gaussian noise, CutOut-style masking to the spectrogram~\cite{cutout}, and slight JPEG compression is added to enhance model performance. Table~\ref{table:aug} summarizes augmentation effects. While CutMix performs well on \textbf{Part I} without perturbation, it negatively impacts scores in the presence of perturbation; others consistently perform well on both \textbf{Part I \& II} data.

\begin{table}
\caption{Effect of Data Augmentation}
\resizebox{8.5cm}{!}{
\label{table:aug}
\begin{tabular}{|c|c|c|c|c|} 
\hline
\multirow{2}{*}{\diagbox{Augmentation}{Score}} & \multicolumn{2}{c|}{Part1} & \multicolumn{2}{c|}{Part2} \\ 
\cline{2-5}
 & CV (F1)  & LB (Acc) & CV (F1) & LB (Acc)  \\ 
\hline
Baseline & 0.962 & 0.948 & 0.937 & 0.934 \\ 
\hline
CutMix~\cite{cutmix} & 0.968 & 0.956 & 0.910 & 0.902 \\ 
\hline
MixUp~\cite{mixup} & 0.965 & 0.951 & 0.948 & 0.940 \\ 
\hline
GaussianNoise & 0.969 & 0.953 & 0.944 & 0.942 \\ 
\hline
JpegCompression & 0.962 & 0.952 & 0.940 & 0.938 \\ 
\hline
Time-Frequency Mask & 0.965 & 0.953 & 0.950 & 0.942 \\ 
\hline
\textbf{Best} & \textbf{0.971} & \textbf{0.962} & \textbf{0.958} & \textbf{0.957} \\
\hline
\end{tabular}
}
\end{table}

\subsubsection{Effect of semi-supervised training}
The pseudo test labels, generated by trained models, contribute to a more robust learning sample space, allowing models to adapt to unknown distributions. In this instance, the pseudo-labels are generated from high-performing models based on the metrics used in evaluation. As a result, despite the possibility of an increased bias towards the training labels, the labels still provide a significant contribution to model training, as observed in Table~\ref{table:method}.

\subsubsection{Effect of ensemble}
The class-wise probabilities from multiple models are averaged to derive the final prediction, enabling the utilization of diverse model insights. As shown in Table~\ref{table:backbone} and Table~\ref{table:method}, it is evident that ensembling significantly improved both CV and LB performance for \textbf{Part I} and \textbf{Part II}. Particularly in \textbf{Part II}, the ensemble increases the results by nearly 2.5\% in observed metrics

\begin{table}
\centering
\caption{LB Scores of Top3 Teams in IEEE SP Cup 2022}
\label{table:sp22}
\resizebox{8cm}{!}{
\begin{tabular}{|l|l|c|c|c|c|} 
\hline
Data                                                                        & \diagbox{Method}{Metric}    & Acc          & Prc          & Rec          & F1             \\ 
\hline
\multirow{3}{*}{\begin{tabular}[c]{@{}l@{}}Eval 1\\(weak pert.)\end{tabular}} & Std. Proc.                  & 0.97          & 0.97          & 0.96          & 0.97           \\ 
\cline{2-6}
                                                                            & Team IITH                   & 0.96          & 0.96          & 0.95          & 0.96           \\ 
\cline{2-6}
                                                                            & \textbf{Synthesizer (Ours)} & \textbf{0.98} & \textbf{0.99} & \textbf{0.97} & \textbf{0.98}  \\ 
\hline
\multirow{3}{*}{\begin{tabular}[c]{@{}l@{}}Eval 2\\(strong pert.)\end{tabular}}  & Std. Proc.                  & 0.48          & 0.62          & 0.48          & 0.48           \\ 
\cline{2-6}
                                                                            & Team IITH                   & 0.49          & 0.51          & 0.49          & 0.49           \\ 
\cline{2-6}
                                                                            & \textbf{Synthesizer (Ours)} & \textbf{0.61} & \textbf{0.71} & \textbf{0.61} & \textbf{0.63}  \\
\hline
\end{tabular}
}
\end{table}

\subsection{Result on IEEE SP Cup 2022}
The performance of the proposed method is rigorously assessed in the IEEE SP Cup~\cite{spcup23} competition at ICASSP 2022. As illustrated in Table~\ref{table:sp22} (w/ ensemble), the proposed method outperforms other top teams on the leaderboard by a significant margin, with an improvement of 12-13\% on Eval 2 (highly perturbed) and 1-2\% on Eval 1 (weakly perturbed), on accuracy metric. This affirms the effectiveness of the method. Notably, Eval 2 dataset is kept hidden from the participants.

\vspace*{-0.2cm}
\subsection{Comparison with Other Approaches}

In Table~\ref{table:method} (w/o ensemble), it becomes evident that the proposed method surpasses other approaches by a considerable margin, in terms of accuracy and F1 score. This superiority can be attributed to the robustness and generalizability of the proposed Unknown Multi-Class Strategy, semi-supervised training, and network ensembling. These findings provide compelling evidence of the effectiveness of the proposed approach in synthetic speech attribution.

\begin{table}
\centering
\caption{Comparison of Different Methods on Eval 1 Data}
\resizebox{8cm}{!}{%
\label{table:method}
\begin{tabular}{|l|c|c|c|c|} 
\hline
\multirow{2}{*}{\diagbox{\textbf{Method}}{\textbf{Score}}} & \multicolumn{2}{c|}{\textbf{Part1}} & \multicolumn{2}{c|}{\textbf{Part2}} \\ 
\cline{2-5}
& \textbf{CV (F1)} & \textbf{LB (Acc)} & \textbf{CV (F1)} & \textbf{LB (Acc)} \\ 
\hline
Xgboost + RandomForest~\cite{xgboost} & 0.443 & 0.427 & 0.422 & 0.409 \\
\hline
Auto-Encoder~\cite{an2015variational} & 0.586 & 0.522 & 0.549 & 0.510 \\ 
\hline
LSTM~\cite{wollmer2013lstm} & 0.696 & 0.645 & 0.637 & 0.608 \\
\hline
ParalMGC~\cite{neri2022paralmgc} & 0.822 & 0.810 & 0.802 & 0.782 \\
\hline
Confidence Threshold~\cite{data-driven} & 0.892 & 0.875 & 0.808 & 0.790 \\
\hline
CAT + t-SNE~\cite{transformer} & 0.901 & 0.881 & 0.861 & 0.854 \\ 
\hline
One-class SVM~\cite{data-driven} & 0.911 & 0.901 & 0.843 & 0.820 \\
\hline
\textbf{Proposed} & \textbf{0.971} & \textbf{0.962} & \textbf{0.958} & \textbf{0.957} \\
\hline
\end{tabular}}
\end{table}

\section{Conclusion}
In this article, a solution for synthetic speech attribution is presented: a semi-supervised multi-class convolutional neural network ensemble-based approach that employs a multi-class strategy with a dedicated unknown class for unidentified algorithms. Its semi-supervised nature ensures effective handling of unknown data distribution whereas the ensemble network enhances detector robustness by incorporating diverse features from different models. Extensive investigation demonstrates its remarkable effectiveness in synthetic speech attribution, notably in the evaluation datasets. It stands as a promising candidate for state-of-the-art synthetic speech attribution, addressing forensic concerns linked to malicious synthetic speech use.

\section{Acknowledgment}
The authors thank IEEE Signal Processing Society, ISPL at Politecnico di Milano (Italy), and MISL at Drexel University (USA) for hosting IEEE SP Cup at ICASSP 2022, which inspired this work.

\printbibliography

\end{document}